\documentclass[10pt,conference,compsocconf]{IEEEtran}

\usepackage{times}

\usepackage{caption}
\usepackage{cite}
\usepackage{url}
\usepackage{times}
\usepackage{xcolor}
\usepackage{amsmath}
\usepackage{tikz}
\usetikzlibrary{shapes,calc,arrows}
\usepackage{graphicx}
\usepackage[english]{babel}
\usepackage{url}

\usepackage{flushend}
\usepackage{orcidlink}
\usepackage{url}

\usepackage{breakurl}
\usepackage{hyperref}
\hypersetup{hidelinks,unicode,breaklinks}

\usepackage[%
babel=true, 
expansion=alltext,
protrusion=alltext-nott, 
nopatch=eqnum, 
final 
]{microtype}

\usepackage{ifpdf}
\ifpdf
\pdfoutput=1 
\pdftrue
\fi

\makeatletter
\def\ps@IEEEtitlepagestyle{
	\def\@oddfoot{\mycopyrightnotice}
	\def\@evenfoot{}
}
\def\mycopyrightnotice{
	{\footnotesize
		\begin{minipage}{0.8\textwidth}
			\centering
			Please cite as: Andreas Aßmuth, Robert Duncan, Simon Liebl, and Matthias Söllner, ``A Secure and Privacy-Friendly Logging Scheme,'' in \emph{Proc of the 12th International Conference on Cloud Computing, GRIDs, and Virtualization (Cloud~Computing~2021), Porto, Portugal}, April 2021.
		\end{minipage}
	}
}
\makeatother
\makeatletter
\let\blx@rerun@biber\relax
\makeatother

\DeclareRobustCommand*{\IEEEauthorrefmark}[1]{%
	\raisebox{0pt}[0pt][0pt]{\textsuperscript{\footnotesize #1}}%
}

\begin{document}
\pagenumbering{gobble}

\title{\textbf{\Large A Secure and Privacy-Friendly Logging Scheme}\\[0.2ex]}

\author{%
	\IEEEauthorblockN{~\\[-0.4ex]\large Andreas A\ss muth\IEEEauthorrefmark{1}\,\orcidlink{0009-0002-2081-2455}, Robert Duncan\IEEEauthorrefmark{3}, Simon Liebl\IEEEauthorrefmark{1,2}\,\orcidlink{0000-0003-1311-4202}, and Matthias S\"ollner\IEEEauthorrefmark{1}\\[0.2ex]\normalsize}
	\IEEEauthorblockA{\IEEEauthorrefmark{1}Technical University of Applied Sciences OTH Amberg-Weiden\\Amberg, Germany\\Email: {\tt \{a.assmuth $|$ s.liebl $|$ m.soellner\}@oth-aw.de}\\[1mm]}
	\IEEEauthorblockA{\IEEEauthorrefmark{2}PhD Student at Abertay University, Dundee, UK\\[1mm]}
	\IEEEauthorblockA{\IEEEauthorrefmark{3}University of Aberdeen\\Aberdeen, United Kingdom\\Email: {\tt robert.duncan@abdn.ac.uk}\\[0.75ex]}
	
}

\maketitle

\begin{abstract}
	Finding a robust security mechanism for audit trail logging has long been a poorly satisfied goal. There are many reasons for this. The most significant of these is that the audit trail is a highly sought after goal of attackers to ensure that they do not get caught. Thus they have an incredibly strong incentive to prevent companies from succeeding in this worthy aim. Regulation, such as the European Union General Data Protection Regulation, has brought a strong incentive for companies to achieve success in this area due to the punitive level of fines that can now be levied in the event of a successful breach by an attacker. We seek to resolve this issue through the use of an encrypted audit trail process that saves encrypted records to a true immutable database, which can ensure audit trail records are permanently retained in encrypted form, with no possibility of the records being compromised. This ensures compliance with the General Data Protection Regulation can be achieved.
\end{abstract}


\IEEEpeerreviewmaketitle

\section{Introduction}
Today, we all are used to authenticate ourselves in order to access systems and services we use in our everyday life. Authentication can be viewed from two different perspectives. For ourselves and especially for our private use, authentication ensures that no one else can access our data. For the system or service provider, authentication is used to distinguish between users. Different users may have different subscriptions for the services and, for example, the service provider is not interested in users using services for free that should be paid for. Additionally, the system provider might get contacted by authorities or law enforcement agencies if illegal actions have occurred involving their services. In this case, authentication is used to determine which user is responsible for the illegal actions and which users were not involved at all.\par 
At work, it is common practice that several employees share a computer or work with one industrial machine. And of course, Cloud-based services and applications allow multiple employees to work on a project collaboratively. In all these systems, it is important to identify and authenticate the current user in order to grant him or her the appropriate privileges. In order to trace who, for example, has processed an order, the digital identity of the employee is saved and logged. In the event of a severe mistake, a negligent operation on an expensive machine or illegal actions, the company wants to ensure to be able to track down the responsible employee. And if we additionally take compliance to security and privacy regulations into account, we must consider so called malicious insider actions as well. These are incidents that were deliberately caused by (former) employees who want to stain the companies reputation or damage the company's equipment. According to a study by the Ponemon Institute, more than $70\,\%$ of participating companies have had more than ten insider-related incidents within a year \cite{ponemon}.\par 
Features and services that are used to track individual actions to single employees can be viewed critically because such measures can violate the privacy of employees. One such example is the so called ``productivity score'' \cite{msproductivity}, which has raised much criticism and was condemned by the press as a means for workplace surveillance \cite{msproductivityguardian}\negmedspace\cite{msproductivitysz}. But even without such services, permanent monitoring employees may be used to assess their productivity. Therefore, especially companies with a strong workers' council are looking for other solutions. Finding such solutions is often also in the company's executive's interest because some companies have been fined in recent years for violations of the General Data Protection Regulation (GDPR). For example, in 2020, the Hamburg Commissioner for Data Protection and Freedom of Information fined H~\&~M (Hennes \& Mauritz) \texteuro\,35.3~million for data protection violations of employees' personal data. The company recorded a considerable amount of highly personal data about their employees' vacation experiences, but also symptoms of illness and diagnoses. In addition, some supervisors acquired a broad knowledge of their employees' private lives through personal and floor talks, ranging from rather harmless details to family issues and religious beliefs. Some of this knowledge was recorded, digitally stored and partly readable by up to 50 other managers throughout the company. The recordings were sometimes made with a high level of detail and recorded over greater periods of time documenting the development of these issues. This practice only came to light when the data became accessible company wide following a misconfiguration error, following which the regulator became involved \cite{For2020}.\par 
Finding a solution to this problem, to be capable of tracking down individuals without violating their privacy, is not trivial. Going back to shady practices some companies used before they discovered the necessity of being able to track down an employees actions if needed, is definitely no suitable solution. Imagine a manufacturing company assigns a group of employees to machine. This would allow to assess the whole team, but would not violate the privacy of individual employees. But in case of mistakes, illegal actions, etc., the company then would not be capable of tracking down the responsible employee. In practice, this approach leads to additional drawbacks concerning security. In order to facilitate work, in this approach such groups of employees very often use only a single, usually rather weak password that is easy to remember (or may even be found on a sticker right at the machine) instead of having strong individual passwords. In case of sabotage by an employee, the responsible person cannot be determined because he or she does not even have to belong to that group of employees.\par 
This paper is structured as follows: in Section~II, we describe possible logging strategies before we address security and privacy challenges in Section~III. In Section~IV, we present our solution for a secure and privacy-friendly logging scheme and further ideas, how our solution can be modified in order to fulfil special or additional requirements. We conclude in Section~V with an outlook on future work.

\section{Logging Strategies}

Logging is usually carried out for the purpose of providing an audit trail of all activities involved in running the system. This is a practice that has long been carried out in the accounting profession to ensure a robust mechanism exists such that in the event of a disaster, the audit trail may be used to restore the accounting records in order to reconstitute the accounts of the organisation. Of course, once logging for this purpose started to be carried out in electronic systems, smart attackers realised that due to inherent weaknesses in database systems, by attacking the audit trail, it was possible to remove evidence of their incursion into the system by deleting or modifying the audit trail records.

While a number of early database systems offered an immutable database option, there were a number of challenges that needed to be overcome. First, the immutable database could not use key fields, meaning retrieval or analysis of the contents of the database would be both cumbersome and slow. Second, and perhaps more importantly, there was nothing to prevent the entire database from being deleted once the attacker gained entry and has escalated sufficient privileges.

This meant that the use of traditional database systems would not be sufficient to achieve our requirements to retain a secure audit trail through logging. This brought about the need to find an alternative immutable database solution instead. One option would have been to use blockchain, which provides the core security for crypto-currencies. However, there is a potential significant overhead in going down this route. The public blockchain relies on thousands of nodes, which are required to perform extensive cryptographic algorithmic computations to ensure a proper consensus of the contents of the blockchain can be guaranteed, but this brings a huge overhead to the equation, since those who perform the cryptographic tasks are looking for a reward for the considerable efforts they provide, meaning considerable extra costs of operation, along with a lesser level of performance due to the huge redundancy on offer.

The alternative solution here would be for the corporate to use a private blockchain, but this also brings challenges. This private blockchain would be provisioned by the corporate, but now their challenge would be to find a balance between choosing the minimal level of blockchain redundancy to improve latency, against being able to retain a sufficient number of nodes securely enough to retain full control over the contents of the blockchain.

However, in 2020, a new start company introduced immudb \cite{immudb2021}, a lightweight, high-speed immutable database that is specifically designed to complement existing transactional database systems. It is tailor made to track changes in the main database system and to then record these transactions, or logs, in the tamperproof immudb.  The immudb system gives you the same cryptographic verification of the integrity of data written with SHA-256 like classic blockchain without the cost and complexity associated with blockchains today. This means that unlike traditional transaction logs, which are very hard to scale, immudb is extremely fast, scalable, robust and open source, making it ideal to incorporate for this purpose. For further details on the immutable storage we refer to \cite{immudbpaper}.

\section{Security and Privacy Challenges}

Security and privacy challenges in this area are not new. In 1999, Schneier and Kelsey \cite{schneier1999secure} set out to secure the collection of sensitive logs using encryption, to ensure that forensic records could be maintained in the event of a cyber breach. Some five years later, Waters et al. \cite{waters2004building}, realised that system logs were becoming a prime attack vector for attackers, who were seeking to cover their trail after successfully breaking in to computer systems. The authors felt that improved searchability would be an asset in dealing with a subsequent forensic examination, and they sought to provide a rapid search function to interrogate this encrypted data. Further, they implemented an audit log for database queries using hash chains for integrity protection and identity-based encryption with extracted keywords to enable better searching. Over a decade later, Syta et al. \cite{syta2016keeping}, felt that such was the interest of attackers in this area that further strengthening of systems would be necessary to ensure proper protection could be achieved. The authors attempted to allow a considerable increase in scale, as well as the development of multi-signatures to provide further protection. Their system is claimed to protect against man-in-the-middle attacks.

\section{Example Logging Approach}
The logging scheme we propose consists of two basic components: (a)~an appropriate secret sharing scheme and (b)~an immutable storage. Readers who are already familiar with immutable storage and secret sharing schemes may want to skip the according subsections.

\subsection{Immutable Storage}
The reason we seek to use immutable storage is to ensure that we can only ever add new records to the database. We are not ever allowed to modify or delete records. This will allow us to create entries of permanent record with which to store any information related to the authentication of employees. This will prevent any party from interfering with any entry of permanent record, ensuring we are able to retain permanency of all such transactions. This will provide an audit trail of all transactions relating to employees. Regardless of whether any attack comes from an external source, or from a malicious inside party, they will not be able to alter any of these records. The data that is stored in this immutable storage is encrypted in order to fulfil the demanded privacy constraints. Since is not possible to tamper with the data stored in this immutable storage, even gaining access to the data will not reveal any interesting details to the attacker. 

\subsection{Secret Sharing}
The idea of secret sharing is as follows: some data $D$ is divided into $n$ pieces, $D_1,\dots,D_n$, in such a way that $D$ can be reconstructed of any $k<n$ pieces $D_i$. Additionally, it is ensured that the knowledge of $k-1$ or less pieces $D_i$ is not sufficient to reconstruct $D$. In this case, the reconstruction ends up with a completely undetermined set of bits. Adi Shamir proposed a secret sharing scheme in 1979 that is based upon polynomial interpolation~\cite{shamir}. To emphasise the importance of the two integers $n$ and $k$, Shamir named such a secret sharing scheme a ``$(k,n)$ threshold scheme''. Following Shamir's blueprint, $D$ is associated with an integer smaller than some prime number $p>D$. For $k$ points $(x_i,y_i)\in \operatorname{GF}(p)\times\operatorname{GF}(p)$, $i=1,\dots,k$, with distinct coordinates $x_i$, there exists one and only one polynomial $q$ of degree $k-1$, such that $q(x_i)=y_i$ holds for all $i=1,\dots,k$. For the polynomial
$$ q(x) = \sum_{i=0}^{k-1} a_ix^i $$
the coefficients $a_1$ to $a_{k-1}$ are chosen randomly and the coefficient $a_0$ is used to store $D$, so $a_0=D$. In order to obtain the $n$ different pieces $D_1,\dots,D_n$, the function values of the indices are computed:
$$ D_i = q(i),\quad i=1,\dots,n. $$
From any subset of $k$ elements $D_i$, the coefficients $a_i$ can be computed, provided that their identifying indices are known. After the polynomial $q$ has been revealed, the reconstruction of the data $D$ is achieved by computing $q(0)=a_0=D$. If Shamir's secret sharing scheme is intended to be used, the first step is to specify $k$, the number of pieces needed for the reconstruction of $D$. The total number of pieces then is $n=2k-1$. As pointed out before,
$$ k = \left\lceil\frac{n+1}{2}\right\rceil $$
or more pieces $D_i$ allow the reconstruction of $D$, whereas less than $k$,
$$ k-1 = \left\lfloor\frac{n}{2}\right\rfloor $$
are not sufficient.\par 
Blakley's solution to the secret sharing problem is based on finite geometry~\cite{blakley}. He suggested to encode the secret as a coordinate of a point in a $k$-dimensional space. The basic idea of Blakley's secret sharing scheme is that any $k$ nonparallel $(k-1)$-dimensional hyperplanes intersect at a specific point which in this case contains the secret. In order to generate the secret shares, $n$ hyperplane equations are computed using the intersection coordinates and additional random numbers modulo a prime number $p$. Any $k$ or more out of these $n$ hyperplane equations may be used to construct a system of linear equations that can easily be solved in order to obtain the secret provided that the determinant of the coefficient matrix formed of the given hyperplane equations is nonzero modulo~$p$.
\par 
For these two traditional secret sharing schemes, the shares are at least of the same size as the secret itself. The authors of this paper have successfully used secret sharing schemes before, e.g., in order to store log files of the nodes of a Cloud system in a decentralised way~\cite{cloud2017,journal2018,bafa}. For applications like this, that require secret sharing schemes to be applied to large amounts of data, this is probably unfavourable. In this case, the work of Krawczyk should be helpful who found out that if the notion of secrecy is reduced to computational instead of information theoretic secrecy, a remarkable amount of space and communication can be reduced \cite{krawczyk}. But as the next subsection is going to clarify, this is not a problem for the proposed logging scheme, because the secret sharing scheme is used to secure only a small amount of data, namely the private key of a public key encryption scheme.

\subsection{Proposed Logging Scheme}
On the basis of the two core components, immutable storage and secret sharing, we describe our solution to the problem in this subsection. Our solution can be applied to a single company site, but also to multiple sites of a (larger) company, which are located in different countries and interconnected using a Cloud service.\par 
To achieve maximum security, all persons must authenticate themselves using individual accounts on the system. Preferably, two-factor authentication should be used. The information, who logged into the system at which time, must be stored encrypted in order to prevent unauthorised personnel from reading this sensitive information. Since we have a system for a whole company in mind, it seems plausible to assume there are several computers or machines that all need to be in the logging system because employees log into all of these computers and machines. All of these devices must be capable of encrypting their logging information and, therefore, need an encryption key to be stored on each device. For our logging system, we choose a public-key encryption scheme, so the encryption key may be stored on all of these devices and is assumed to be publicly known. An encryption scheme with symmetric keys that uses the same key for encryption and decryption is not suitable in this case because of the necessity to have the encryption key stored on a large number of devices. This secret key might fall into the hands of an unauthorised person, e.g., from a single unsecured computer, and this person would then also be capable of reading all the sensitive logging information. Thus, in order to be able to read the sensitive information, the corresponding private key is required for decryption. This key is not stored on these computers and machines because there is no need to decrypt the data locally. The private key is divided into a number of parts, e.g., into three parts: one part for the employer, one part for the workers' council representing the employees and one part for law enforcement authorities. It might be sensible to have more or other groups, therefore, we do not stick to this example but just count these different groups, which all get a part of the secret key. It must be stressed that all of these parts are needed to reconstruct the private key in order to decrypt the encrypted logging data. So all of the groups must agree and combine their private key parts (AND operation).
\begin{figure}[htbp]
	\centering%
	\begin{tikzpicture}
		\footnotesize
		\node[draw, thick, fill=black!10] at (0, 0) {Logging System};
		\node[rotate=-20] at (1.45, 0) {\includegraphics[height=8mm]{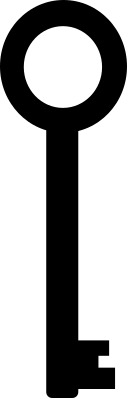}};
		\node[anchor=west] at (1.5, 0) {public key (encryption)};
		\node[rotate=-20] at (0, -0.75) {\includegraphics[height=8mm]{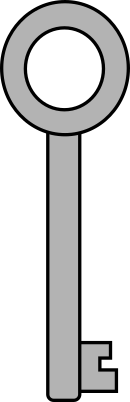}};
		\node[anchor=west] at (0.05, -0.75) {private key (decryption)};
		\draw[thick, -latex] (-0.25, -1) -- ++ (-1, -0.5);
		\draw[thick, -latex] (0.05, -1.1) -- ++ (0.25, -0.5);
		\draw[thick, -latex] (0.15, -1) -- ++ (1.7, -0.5);
		\node at (-0.35, -1.35) {\scriptsize\&};
		\node at (0.6, -1.35) {\scriptsize\&};
		\node[anchor=north east] at (-1.25, -1.55) {\includegraphics[height=0.75cm]{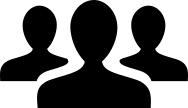}};
		\node[anchor=north] at (0.3, -1.55) {\includegraphics[height=0.75cm]{group.png}};
		\node[anchor=north west] at (1.85, -1.55) {\includegraphics[height=0.75cm]{group.png}};
		\node at (-2, -2.6) {Group 1};
		\node at (0.3, -2.6) {Group 2};
		\node at (2.6, -2.6) {Group 3};
		\draw[thick, -latex] (-2, -2.8) -- ++(.2, -2.4);
	\draw[thick, -latex] (0.3, -2.8) -- ++(.2, -2.4);
	\draw[thick, -latex] (2.6, -2.8) -- ++(.2, -2.4);
	
	\draw[thick, -latex] (-2, -2.8) -- ++(.4, -2.5);
	\draw[thick, -latex] (0.3, -2.8) -- ++(.4, -2.5);
	\draw[thick, -latex] (2.6, -2.8) -- ++(.4, -2.5);
	
	\draw[thick, -latex] (-2, -2.8) -- ++(-.2, -2.4);
	\draw[thick, -latex] (0.3, -2.8) -- ++(-.2, -2.4);
	\draw[thick, -latex] (2.6, -2.8) -- ++(-.2, -2.4);
	
	\draw[thick, -latex] (-2, -2.8) -- ++(-.4, -2.5);
	\draw[thick, -latex] (0.3, -2.8) -- ++(-.4, -2.5);
	\draw[thick, -latex] (2.6, -2.8) -- ++(-.4, -2.5);
	
	\node[draw, thick, fill=white] at (0.3, -4.1) {\begin{minipage}{1.8cm}
		Secret Sharing: split the shared secret 
		\end{minipage}};
	\node[draw, thick, fill=white] at (-2, -4.1) {\begin{minipage}{1.8cm}
		Secret Sharing: split the shared secret 
		\end{minipage}};
	\node[draw, thick, fill=white] at (2.6, -4.1) {\begin{minipage}{1.8cm}
		Secret Sharing: split the shared secret 
		\end{minipage}};
	\node at (-2.35, -5.7) {\includegraphics[height=0.6cm]{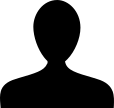}}; 
	\node at (-1.65, -5.7) {\includegraphics[height=0.6cm]{person.png}};
	\node at (-2.35, -6.35) {\includegraphics[height=0.6cm]{person.png}};
	\node at (-1.65, -6.35) {\includegraphics[height=0.6cm]{person.png}};
	\node at (-0.05, -5.7) {\includegraphics[height=0.6cm]{person.png}}; 
	\node at (0.65, -5.7) {\includegraphics[height=0.6cm]{person.png}};
	\node at (-0.05, -6.35) {\includegraphics[height=0.6cm]{person.png}};
	\node at (0.65, -6.35) {\includegraphics[height=0.6cm]{person.png}};
	\node at (2.25, -5.7) {\includegraphics[height=0.6cm]{person.png}}; 
	\node at (2.95, -5.7) {\includegraphics[height=0.6cm]{person.png}};
	\node at (2.25, -6.35) {\includegraphics[height=0.6cm]{person.png}};
	\node at (2.95, -6.35) {\includegraphics[height=0.6cm]{person.png}};
	
	\end{tikzpicture}
	\setlength{\belowcaptionskip}{-12pt}
	\caption{Distribution of the private key among several groups and persons.\\[0.25cm]}
	\label{fig:private-key}
\end{figure}
Figure~\ref{fig:private-key} depicts the fragmentation of the private key and the distribution of the parts to three groups. In practice, these parts of the private key would possibly be in possession of one person of the respective group. But this would be quite unfavourable because this makes that single person a high-value target for attacks that aim to get the respective part of the secret key. Additionally, if there is an incident, it must be dealt with instantly. It would be unacceptable if that one person would then be in another shift, ill at home or on leave. That one person might also be threatened or bribed to give access to his or her part of the secret key; or that person might accidentally loose or delete their part of the private key. For all these reasons, it makes sense to divide the secret key part of each group into pieces using a secret sharing scheme. These pieces are then given to $n$~persons of each group and it takes $k$ of them to agree in order to merge the private key part of the group.\par 
To sum up, this logging system supports both: security and privacy. Employees use their strong individual credentials for authentication. But they must not fear workplace surveillance or an unauthorised assessment of their productivity because their employer is not capable of reading the log files arbitrarily. In case an incident occurs and there is an official investigation, the different groups combine their parts to reconstruct the private key for the decryption of the log files. For each group, access to the respective part of the secret key is granted if a majority of group members ($k$ out of $n$) agree.

\subsection{Adaptability to Certain Scenarios}

\begin{figure}[htbp]
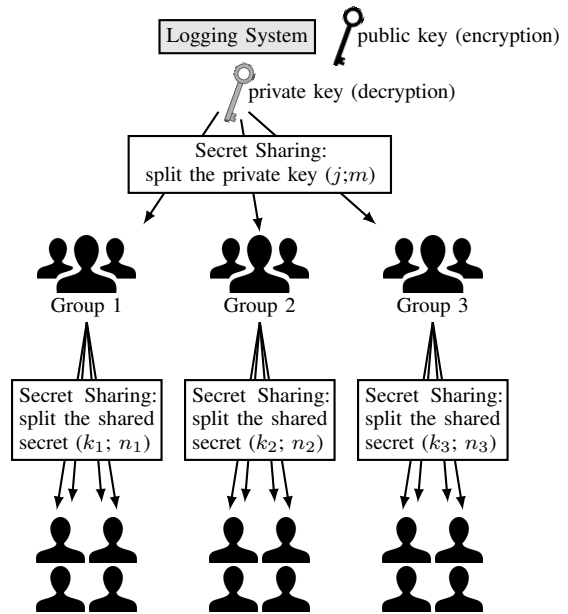

	\centering%
	\begin{tikzpicture}
	\footnotesize
	\node[draw, thick, fill=black!10] at (0, 0) {Logging System};
	\node[rotate=-20] at (1.45, 0) {\includegraphics[height=8mm]{public-key.png}};
	\node[anchor=west] at (1.5, 0) {public key (encryption)};
	\node[rotate=-20] at (0, -0.75) {\includegraphics[height=8mm]{secret-key.png}};
	\node[anchor=west] at (0.05, -0.75) {private key (decryption)};
	\draw[thick, -latex] (-0.25, -1) -- ++ (-1, -1.5);
	\draw[thick, -latex] (0.05, -1.1) -- ++ (0.25, -1.5);
	\draw[thick, -latex] (0.15, -1) -- ++ (1.7, -1.5);
		\node[draw, thick, fill=white] at (0.3, -1.7) {\begin{minipage}{3.3cm}\begin{center}
			Secret Sharing: \\split the private key ($j$;$m$)
			\end{center}\end{minipage}};
	\node[anchor=north east] at (-1.25, -2.55) {\includegraphics[height=0.75cm]{group.png}};
	\node[anchor=north] at (0.3, -2.55) {\includegraphics[height=0.75cm]{group.png}};
	\node[anchor=north west] at (1.85, -2.55) {\includegraphics[height=0.75cm]{group.png}};
	\node at (-2, -3.6) {Group 1};
	\node at (0.3, -3.6) {Group 2};
	\node at (2.6, -3.6) {Group 3};
	\draw[thick, -latex] (-2, -3.8) -- ++(.2, -2.4);
	\draw[thick, -latex] (0.3, -3.8) -- ++(.2, -2.4);
	\draw[thick, -latex] (2.6, -3.8) -- ++(.2, -2.4);

	\draw[thick, -latex] (-2, -3.8) -- ++(.4, -2.5);
    \draw[thick, -latex] (0.3, -3.8) -- ++(.4, -2.5);
    \draw[thick, -latex] (2.6, -3.8) -- ++(.4, -2.5);

	\draw[thick, -latex] (-2, -3.8) -- ++(-.2, -2.4);
\draw[thick, -latex] (0.3, -3.8) -- ++(-.2, -2.4);
\draw[thick, -latex] (2.6, -3.8) -- ++(-.2, -2.4);

\draw[thick, -latex] (-2, -3.8) -- ++(-.4, -2.5);
\draw[thick, -latex] (0.3, -3.8) -- ++(-.4, -2.5);
\draw[thick, -latex] (2.6, -3.8) -- ++(-.4, -2.5);

	\node[draw, thick, fill=white] at (0.3, -5.1) {\begin{minipage}{1.8cm}
		Secret Sharing: split the shared secret ($k_2$; $n_2$)
		\end{minipage}};
		\node[draw, thick, fill=white] at (-2, -5.1) {\begin{minipage}{1.8cm}
		Secret Sharing: split the shared secret ($k_1$; $n_1$)
		\end{minipage}};
		\node[draw, thick, fill=white] at (2.6, -5.1) {\begin{minipage}{1.8cm}
		Secret Sharing: split the shared secret ($k_3$; $n_3$)
		\end{minipage}};
	\node at (-2.35, -6.7) {\includegraphics[height=0.6cm]{person.png}}; 
	\node at (-1.65, -6.7) {\includegraphics[height=0.6cm]{person.png}};
	\node at (-2.35, -7.35) {\includegraphics[height=0.6cm]{person.png}};
	\node at (-1.65, -7.35) {\includegraphics[height=0.6cm]{person.png}};
	\node at (-0.05, -6.7) {\includegraphics[height=0.6cm]{person.png}}; 
	\node at (0.65, -6.7) {\includegraphics[height=0.6cm]{person.png}};
	\node at (-0.05, -7.35) {\includegraphics[height=0.6cm]{person.png}};
	\node at (0.65, -7.35) {\includegraphics[height=0.6cm]{person.png}};
	\node at (2.25, -6.7) {\includegraphics[height=0.6cm]{person.png}}; 
	\node at (2.95, -6.7) {\includegraphics[height=0.6cm]{person.png}};
	\node at (2.25, -7.35) {\includegraphics[height=0.6cm]{person.png}};
	\node at (2.95, -7.35) {\includegraphics[height=0.6cm]{person.png}};

	\end{tikzpicture}
\caption{Adaption of the system: Secret Sharing among groups.}
\label{fig:private-key2}
\end{figure}

The presented scheme proposes that the different groups have to combine their parts of the private key to get the full private key and gain access. As the parts of the private key are combined with an AND operation, all groups have to  contribute to gain access. On the other side scenarios might be interesting and desirable, where it would be sufficient when only $j$ out of $m$ groups come together to combine their keys in order to  access the data. For this purpose the logging scheme can be adapted to share the private key among the groups also with the same secret sharing principle and inside a group the shared part can be shared with this scheme as proposed above (cf. Figure~\ref{fig:private-key2}).

By using this adapted scheme it is possible to gain access to the secret, if only $j$ out of $m$ groups come together to combine the private key and in every group it would take only $k$ out of $n$ members of this group to agree to reconstruct their part of the shared secret. As it is only a matter of design, how many group members are needed to reconstruct their partial secret, the scheme can be adapted very flexibly to different scenarios: Each group $g$ can have its own $k_g$ and $n_g$. So, for example, group~1 has size $n_1$ and $k_1$ members of this group have to agree, group~2 may be much larger ($n_2>n_1$) but fewer members ($k_2$) are needed in order to reconstruct their group's part of the secret key, and so on. 

\begin{figure}[htbp]
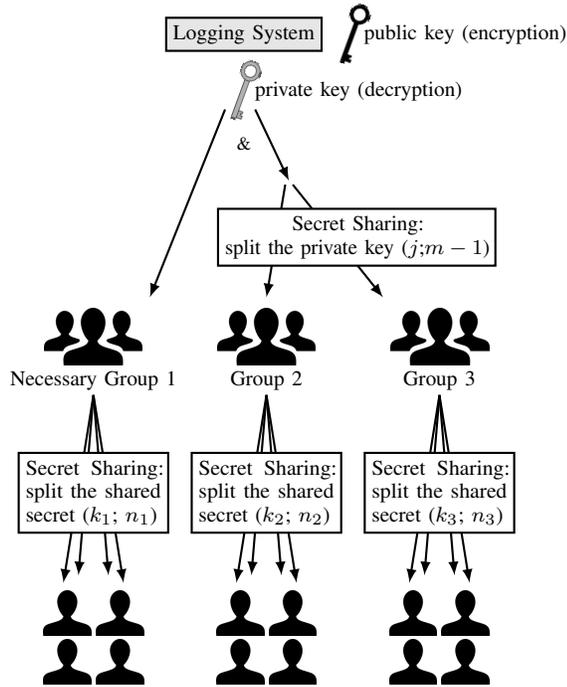

	\centering%
	\begin{tikzpicture}
	\footnotesize
	\node[draw, thick, fill=black!10] at (0, 1) {Logging System};
	\node[rotate=-20] at (1.45, 1) {\includegraphics[height=8mm]{public-key.png}};
	\node[anchor=west] at (1.5, 1) {public key (encryption)};
	\node[rotate=-20] at (0, 0.25) {\includegraphics[height=8mm]{secret-key.png}};
	\node[anchor=west] at (0.05, 0.25) {private key (decryption)};
	\draw[thick, -latex] (-0.25, -0) -- ++ (-1, -2.5);
		\draw[thick, -latex] (+0.15, -0) -- ++ (.45, -.95);
	\draw[thick, -latex] (0.55, -1) -- ++ (-0.25, -1.5);
	\draw[thick, -latex] (0.65, -1) -- ++ (1.2, -1.5);
		\node at (-0, -0.45) {\scriptsize\&};
	\node[draw, thick, fill=white] at (1.5, -1.7) {\begin{minipage}{3.5cm}\begin{center}
		Secret Sharing: \\split the private key ($j$;$m-1$)
		\end{center}\end{minipage}};
	\node[anchor=north east] at (-1.25, -2.55) {\includegraphics[height=0.75cm]{group.png}};
	\node[anchor=north] at (0.3, -2.55) {\includegraphics[height=0.75cm]{group.png}};
	\node[anchor=north west] at (1.85, -2.55) {\includegraphics[height=0.75cm]{group.png}};
	\node at (-2, -3.6) {Necessary Group 1};
	\node at (0.3, -3.6) {Group 2};
	\node at (2.6, -3.6) {Group 3};
	\draw[thick, -latex] (-2, -3.8) -- ++(.2, -2.4);
	\draw[thick, -latex] (0.3, -3.8) -- ++(.2, -2.4);
	\draw[thick, -latex] (2.6, -3.8) -- ++(.2, -2.4);
	
	\draw[thick, -latex] (-2, -3.8) -- ++(.4, -2.5);
	\draw[thick, -latex] (0.3, -3.8) -- ++(.4, -2.5);
	\draw[thick, -latex] (2.6, -3.8) -- ++(.4, -2.5);
	
	\draw[thick, -latex] (-2, -3.8) -- ++(-.2, -2.4);
	\draw[thick, -latex] (0.3, -3.8) -- ++(-.2, -2.4);
	\draw[thick, -latex] (2.6, -3.8) -- ++(-.2, -2.4);
	
	\draw[thick, -latex] (-2, -3.8) -- ++(-.4, -2.5);
	\draw[thick, -latex] (0.3, -3.8) -- ++(-.4, -2.5);
	\draw[thick, -latex] (2.6, -3.8) -- ++(-.4, -2.5);
	
	\node[draw, thick, fill=white] at (0.3, -5.1) {\begin{minipage}{1.8cm}
		Secret Sharing: split the shared secret ($k_2$; $n_2$)
		\end{minipage}};
	\node[draw, thick, fill=white] at (-2, -5.1) {\begin{minipage}{1.8cm}
		Secret Sharing: split the shared secret ($k_1$; $n_1$)
		\end{minipage}};
	\node[draw, thick, fill=white] at (2.6, -5.1) {\begin{minipage}{1.8cm}
		Secret Sharing: split the shared secret ($k_3$; $n_3$)
		\end{minipage}};
	\node at (-2.35, -6.7) {\includegraphics[height=0.6cm]{person.png}}; 
	\node at (-1.65, -6.7) {\includegraphics[height=0.6cm]{person.png}};
	\node at (-2.35, -7.35) {\includegraphics[height=0.6cm]{person.png}};
	\node at (-1.65, -7.35) {\includegraphics[height=0.6cm]{person.png}};
	\node at (-0.05, -6.7) {\includegraphics[height=0.6cm]{person.png}}; 
	\node at (0.65, -6.7) {\includegraphics[height=0.6cm]{person.png}};
	\node at (-0.05, -7.35) {\includegraphics[height=0.6cm]{person.png}};
	\node at (0.65, -7.35) {\includegraphics[height=0.6cm]{person.png}};
	\node at (2.25, -6.7) {\includegraphics[height=0.6cm]{person.png}}; 
	\node at (2.95, -6.7) {\includegraphics[height=0.6cm]{person.png}};
	\node at (2.25, -7.35) {\includegraphics[height=0.6cm]{person.png}};
	\node at (2.95, -7.35) {\includegraphics[height=0.6cm]{person.png}};

	\end{tikzpicture}
	\setlength{\belowcaptionskip}{-8pt}
	\caption{Further adaption of the system: Secret Sharing among groups and making one group necessary.}
	\label{fig:private-key3}
\end{figure}

Furthermore, it is also  possible to include one or more groups, which have to contribute necessarily (e.g., group~1 in Figure~\ref{fig:private-key3}), because  it is possible to combine AND operation and secret sharing on the group level, too. This means the private key is first split in two (or more to include more necessary groups) parts, which have to be combined again with AND operation later. One of this parts can then be distributed with secret sharing, the other parts are only shared within the necessary groups.

\section{Conclusion and Future Work}

As a first step, we have developed a highly secure logging approach for logging events connected with employees within the organisation. The logging data is captured and fully encrypted to ensure full compliance with the GDPR for any PII relating to employees of the organisation, since the data cannot be identified by anyone other than the duly authorised users of the system. We have demonstrated that this approach can deliver exactly the high security level of employee privacy which we were seeking.\par 
Our next step will be to plan for the implementation of a proof-of-concept solution. As part of this process, we would test the outcome and performance of the system using differing secret sharing schemes to ensure we can deliver the most effective and powerful solution. However, we would also consider the possibility of investigating the development of how a verifiable secret sharing solution might further improve our suggested scheme.\par 
Once we have reached that stage, we would seek to carry out an investigation into possible practical issues and endeavour to recognise any remaining problems with this work. We consider there may be the possibility of a collaboration between the two universities, OTH Amberg-Weiden and the University of Aberdeen.

\end{document}